\documentclass[prd,reprint,nofootinbib,notitlepage,aps,tightenlines,
preprintnumbers,amsmath,amssymb,showpacs,superscriptaddress]{revtex4-1}

\usepackage[hyperindex,breaklinks]{hyperref}
\usepackage{graphicx,epstopdf,amsmath,amsfonts,amssymb,color}

\def\be{\begin{equation}}
\def\ee{\end{equation}}
\def\ba{\begin{eqnarray}}
\def\ea{\end{eqnarray}}
\def\ge{\mathrel{\raise.3ex\hbox{$>$\kern-.75em\lower1ex\hbox{$\sim$}}}}
\def\la{\mathrel{\raise.3ex\hbox{$<$\kern-.75em\lower1ex\hbox{$\sim$}}}}

\def\simgt{\mathrel{\raise.3ex\hbox{$>$\kern-.75em\lower1ex\hbox{$\sim$}}}}
\def\simlt{\mathrel{\raise.3ex\hbox{$<$\kern-.75em\lower1ex\hbox{$\sim$}}}}

\newcommand{\bi}[1]{\bibitem{#1}}

\newcommand{\nc}{\newcommand}

\nc{\gone}{\bar g_{\pi NN}^{(1)}}
\nc{\gzero}{\bar g_{\pi NN}^{(0)}}
\nc{\al}{\alpha}
\nc{\ga}{\gamma}
\nc{\de}{\delta}
\nc{\ep}{\epsilon}
\nc{\ze}{\zeta}
\nc{\et}{\eta}
\nc{\ka}{\kappa}
\nc{\rh}{\rho}
\nc{\si}{\sigma}
\nc{\ta}{\tau}
\nc{\up}{\upsilon}
\nc{\ph}{\phi}
\nc{\ch}{\chi}
\nc{\ps}{\psi}
\nc{\om}{\omega}
\nc{\Ga}{\Gamma}
\nc{\De}{\Delta}
\nc{\La}{\Lambda}
\nc{\Si}{\Sigma}
\nc{\Up}{\Upsilon}
\nc{\Ph}{\Phi}
\nc{\Ps}{\Psi}
\nc{\Om}{\Omega}
\nc{\ptl}{\partial}
\nc{\del}{\nabla}
\nc{\ov}{\overline}
\nc{\newcaption}[1]{\centerline{\parbox{15cm}{\caption{#1}}}}
\nc{\us}{U(1)$_S$}
\nc{\Rg}{$R_{\gamma\gamma}$}

\def\beq{\begin{equation}}
\def\eeq{\end{equation}}
\def\bmat{\begin{displaymath}}
\def\emat{\end{displaymath}}
\def\bear{\begin{eqnarray}}
\def\eear{\end{eqnarray}}
\def\ba{\begin{eqnarray}}
\def\ea{\end{eqnarray}}
\def\bery{\begin{array}}
\def\ery{\end{array}}
\def\bit{\begin{itemize}}
\def\eit{\end{itemize}}
\def\ben{\begin{enumerate}}
\def\een{\end{enumerate}}
\def\btab{\begin{tabular}}
\def\etab{\end{tabular}}
\def\btbl{\begin{table}}
\def\etbl{\end{table}}
\def\bfig{\begin{figure}[htb]}
\def\efig{\end{figure}}
\def\bpic{\begin{picture}}
\def\epic{\end{picture}}

\def\ga{\mathrel{\raise.3ex\hbox{$>$\kern-.75em\lower1ex\hbox{$\sim$}}}}
\def\la{\mathrel{\raise.3ex\hbox{$<$\kern-.75em\lower1ex\hbox{$\sim$}}}}
\def\gappeq{\mathrel{\rlap {\raise.5ex\hbox{$>$}}
{\lower.5ex\hbox{$\sim$}}}}
\def\lappeq{\mathrel{\rlap{\raise.5ex\hbox{$<$}}
{\lower.5ex\hbox{$\sim$}}}}

\def\gyr{{\rm \, G\kern-0.125em yr}}
\def\mev{{\rm \, Me\kern-0.125em V}}
\def\gev{{\rm \, Ge\kern-0.125em V}}
\def\tev{{\rm \, Te\kern-0.125em V}}

\def\lsim{\mathrel{\rlap{\lower4pt\hbox{\hskip1pt$\sim$}}
    \raise1pt\hbox{$<$}}}                
\def\gsim{\mathrel{\rlap{\lower4pt\hbox{\hskip1pt$\sim$}}
    \raise1pt\hbox{$>$}}}                

\begin{document}
 
\title{Long-range axion forces and hadronic CP violation}

\author{Shohei Okawa}
\affiliation{Department of Physics and Astronomy, University of Victoria, 
Victoria, BC V8P 5C2, Canada}
\affiliation{Departament de F\'isica Qu\`antica i Astrof\'isica, Institut de Ci\`encies del Cosmos (ICCUB),
Universitat de Barcelona, Mart\'i i Franqu\`es 1, E-08028 Barcelona, Spain}

\author{Maxim Pospelov}
\affiliation{School of Physics and Astronomy, University of Minnesota, Minneapolis, MN 55455, USA}
\affiliation{William I. Fine Theoretical Physics Institute, School of Physics and Astronomy,
University of Minnesota, Minneapolis, MN 55455, USA}

\author{Adam Ritz}
\affiliation{Department of Physics and Astronomy, University of Victoria, 
Victoria, BC V8P 5C2, Canada}

\date{November 2021}

\begin{abstract}
Axions and other pseudoscalar fields comprise an interesting class of ultralight dark matter candidates, that may independently play a role in solving the strong $CP$ problem. In the presence of $CP$-violating sources, these pseudoscalar fields can develop a coherent non-derivative coupling to nucleons, $\bar g_{aNN}$, 
thus mediating `mass-mass' and `mass-spin' forces in matter that can be probed experimentally. 
We revisit the non-perturbative generation of these $CP$-odd axion forces, and refine estimates of 
$\bar g_{aNN}$ generated by the EDMs and color EDMs of quarks.  We also revisit the Standard Model contribution to $CP$-odd axion couplings generated by the phase of the Cabibbo-Kobayashi-Maskawa quark mixing matrix. 
\end{abstract}
\maketitle

\section{Introduction}
\label{sec:intro}

Axions have long been considered a compelling dark matter candidate \cite{Preskill:1982cy,Abbott:1982af,Dine:1982ah}, given their independent motivation as a potential resolution of the strong $CP$ problem. In recent years, with increasing constraints on scenarios of thermal relic WIMP dark matter, further attention has been paid more generally to light or ultralight (pseudo)scalar fields as dark matter candidates, or as components of light hidden or dark sectors. Such ultralight bosonic degrees of freedom may escape conventional direct detection and require novel search strategies. 

One such strategy is to consider precision tests for new long-range forces. Pseudoscalar fields naively escape the strongest constraints, since they couple to spin, and thus do not mediate Yukawa-type forces. However, this conclusion isn't directly applicable in the presence of $CP$-violation, as all Lorentz-scalar fields then acquire scalar and pseudoscalar components. Long-range forces mediated by axions are thus suppressed only by the small scale of $CP$-violation. The possibility of detecting axions via this channel was first considered by Moody and Wilczek \cite{MW}, and has been addressed a number of times in the literature. In the presence of $CP$-violation, axions can mediate a (scalar) monopole-monopole potential  $V^{ss}_a(r{\rightarrow}\infty) \propto e^{-m_a r}/r$ and also a spin-dependent monopole-dipole potential $V^{sp}_a(r{\rightarrow}\infty) \propto (\hat\sigma \cdot \hat r) e^{-m_a r}/r$, in addition to the generic spin-dependent dipole-dipole interactions. In this paper, we revisit the problem of computing the scalar (as opposed to pseudoscalar) coupling of axions to nucleons  $\bar g_{aNN}$ induced by hadronic $CP$-violating phases, as the primary input to monopole-monopole ($\propto \bar g_{aNN}^2$) and monopole-dipole ($\propto \bar g_{aNN}$) long-range potentials in matter.

We consider three light quark flavors, transforming the defining axion-gluon Lagrangian 
\be
 {\cal L} = 
 \frac{a}{f_a} \frac{\al_s}{8\pi} G^a_{\mu\nu} \tilde{G}^a_{\mu\nu},
 \label{aGG} 
\ee
to the more convenient form, 
\be
 {\cal L} = 
 - m_* \frac{a}{f_a} (\bar{u} i\gamma_5 u + \bar{d}i\gamma_5 d+\bar{s}i\gamma_5 s) + {\cal O}(a/f_a)^2, \label{La}
\ee
where $m_* = \left(\sum_{u,d,s}m_i^{-1}\right)^{-1}\simeq m_um_d/(m_u+m_d)$. 
Nonpertubative effects, as manifest in the strong breaking of $U(1)_A$ symmetry which leads to $m_{\eta'} \gg m_{\rm octet}$, 
generate the axion mass
\begin{equation}
\label{ma}
m_a^2 = m_\pi^2 \left(\frac{F_\pi}{f_a}\right)^2 \frac{m_*}{m_u+m_d},
\end{equation}
where $F_\pi\simeq 93$\,MeV. 
Throughout this paper, we will adhere to the conventional view of the axion mass, corresponding to the relation (\ref{ma}).

The presence of  $CP$-violation in the theory allows the axion to have scalar couplings, and in particular a scalar axion-nucleon coupling,  $\bar g_{aNN}a\bar{N}N$. We focus on determining a quantitative lower bound on the scale of these couplings due to the CKM $CP$-violating phase, the residual $\theta_{\rm eff}$-parameter, and the quark EDM ($d_q$) and chromo-EDM ($\tilde{d}_q$) sources. 
In other words, we are interested in determining 
\begin{equation}
\bar g_{aNN}= \bar g_{aNN}(\bar \theta, d_i, \tilde d_i, \delta_{CKM}),
\end{equation}
and identifying the hadronic matrix elements controlling the size of these couplings. Relevant past work on this subject includes estimates of $\bar g_{aNN}$ due to $\delta_{CKM}$ \cite{GR}, $\tilde d_i$ \cite{P97,Barbieri:1996vt}, as well as recent literature \cite{Bigazzi:2019hav,Bertolini:2020hjc,DiLuzio:2021jfy} on $\bar g_{aNN}$ induced by $CP$-odd four-quark operators. Astrophysical limits on $\bar g_{aNN}$ have been updated in \cite{OHare:2020wah}.

The rest of this paper is organized as follows. In Section~2 we examine the scalar axion-nucleon coupling induced by several $CP$-odd sources in turn. The results are discussed in Section~3, along with some concluding remarks.

\section{Scalar axion-nucleon couplings}

In the low energy limit, the axion couplings to quarks and gluons result in a mass term that can have its minimum shifted from the field value $a=0$ due to the presence of the $CP$-violating sources. Thus at energies well below the hadronic mass scale, and not too far from the minimum, the axion potential can be written in the form
\begin{equation}
\label{thetaind}
V(a) = \frac12 m_a^2 (a - \langle a \rangle )^2 \equiv \frac12 m_a^2 (a - f_a\theta_{\rm ind})^2.
\end{equation}
Here the expectation value of the axion field $\langle a \rangle$ refers to the vacuum value, {\em i.e.} the value where the vacuum energy is minimized, ${\rm min} [V_{vac}(a) ]= V_{vac}(\langle a \rangle)$. 
The physical axion field, by definition, is the deviation of $a$ field from $\langle a \rangle$,
\begin{equation}
a_{\rm ph} \equiv a - \langle a \rangle;~~a = a_{\rm ph} +f_a \theta_{\rm ind}.
\end{equation}

In this section, we analyze scalar axion-nucleon couplings $g_{aNN}$, defined as the coupling of the nucleon scalar densities to 
$a_{\rm ph} $,
\be
 {\cal L} = - \bar g^{(0)}_{aNN} a_{\rm ph}\bar{N}N - \bar g^{(1)}_{aNN} a_{\rm ph} \bar{N}\tau_3 N,
\ee
where $N= (p,n)$ is the nucleon doublet. The isospin singlet part, $\bar g^{(0)}_{aNN}$ is the most important, 
being enhanced by $A$, the total number of nucleons in the nucleus. For that reason, we will often drop the ``0" superscript. 

It is important to emphasize that we will focus on the {\em non-derivative} coupling of axions to nucleons, that is proportional to the lowest power of $f_a^{-1}$,  $g_{aNN}\propto f_a^{-1}$. This is in contrast to derivative couplings, such as $\bar NN \Box a$, that lead either to contact interactions, or to effects proportional to the square of the axion mass, which are therefore proportional by $f_a^{-3}$. 

We classify the various sources of $CP$-violation in a derivative expansion as follows:
\begin{equation}
{\cal L }_{CP}  =  {\cal L}_{\rm IR} +  {\cal L}_{\rm SM}(\delta) + {\cal L}_{{\rm dim} \geq 5}. \label{LCP}
\end{equation} 
The higher-dimensional terms here are induced by $CP$-odd physics beyond the Standard Model (SM) that preserve the Peccei-Quinn (PQ) symmetry, 
and we consider the lowest dimension operators, 
\begin{eqnarray}
{\cal L}_{{\rm dim} \geq 5} = -\sum_{q=u,d,s} 
\left[\frac{i}{2} {d}_q \bar{q} F_{\mu\nu}\sigma_{\mu\nu} \gamma_5 q \right.
\\
\nonumber \left.
 + \frac{i}{2} \widetilde{d}_q \bar{q} G^a_{\mu\nu}\sigma_{\mu\nu}t^a \gamma_5 q \right]+\cdots
\end{eqnarray} 
where $d_q$ and $\widetilde{d}_q$ stand for the EDMs and color EDMs of light quarks, and the ellipsis  
represents higher-dimensional terms including the Weinberg operator and four-fermion interactions. 
From now on we will use the condensed notation $G\si = g_st^aG^a_{\mu\nu} \si^{\mu\nu}$. 

The source ${\cal L}_{\rm SM}(\delta)$ in (\ref{LCP}) denotes the SM source of $CP$-violation where $\delta = \delta_{CKM}$. From the low energy perspective, integrating out heavy quarks and $W$-bosons can generate CP-violating effects in 
the flavour-conserving channels only at ${\cal O}(G_F^2)$. Moreover, as is well known, the overall $CP$-violating effect in such a channel is necessarily proportional to the small combination of the CKM angles given by the Jarlskog invariant $J\sim 10^{-5}$ \cite{jarlskog}.

Finally, we include a rarely discussed possibility: an additional source of soft PQ symmetry breaking, associated with the lowest dimension axion field operators: 
\begin{equation}
 {\cal L}_{\rm IR}=  {\Lambda}^4_{\rm IR}(a/f_a) +\cdots
\end{equation}
The only effect of such a tadpole operator is to shift the axion minimum and induce the theta term,
\begin{equation}
\label{thetaIR}
\theta_{\rm ind} = \frac{\langle {a}  \rangle}{f_a}  = \frac{{\Lambda}^4_{\rm IR}}{m_\pi^2 F_\pi^2m_*(m_u+m_d)^{-1}}.
\end{equation}
This `infrared' breaking of PQ symmetry, associated for example with an additional nonabelian dark sector, $(\theta'+a/f_a)G'\tilde G'\to {\cal L}_{\rm IR}$, with a low condensation scale ${\Lambda}_{\rm IR}\ll {\Lambda}_{\rm QCD}$ and a new source of $CP$-violation ({\em e.g.} $\theta'$), is the only example where the $\bar g_{aNN} $ coupling is associated {\em solely} with the induced value of theta. In all the other examples considered, the contribution of the induced value of theta is accompanied by direct contributions to $\bar g_{aNN}$. 

To carefully define our terminology, we will make a distinction between the various ``induced" values of $\theta$, which are sometimes confused in the literature. First of all, there are additive tree level and radiative corrections to the theta term, $\theta_{\rm rad}$, from colored massive fields, {\em i.e.} quarks. These corrections can be absorbed into a redefinition of the theta term, and the axion field. In the presence of an axion, they do not lead to any new physical effects, and do not induce $\bar g_{aNN}$. To simplify the expressions, all such corrections are already included in the definition of the axion field in (\ref{aGG}). An important quantity in our discussion is the induced value of $\theta$, that can be identified with the expectation value of the axion field as in Eq.~(\ref{thetaind}). As stated above, this induced value of theta can be the result of non-trivial IR physics, or it can be induced by higher-dimensional operators, as discussed in \cite{review}, generically formulated as follows,
\begin{equation}
\theta_{\rm ind} = - \frac{\int d^4x \langle T[\frac{\alpha_s}{8\pi}G\tilde G(0),{\cal L }_{{\rm dim}\geq 5}(x) ]\rangle_{\rm vac} }
{\int d^4x \langle T[\frac{\alpha_s}{8\pi}G\tilde G(0),\frac{\alpha_s}{8\pi}G\tilde G(0)(x) ]\rangle_{\rm vac}}.
\end{equation}
 Finally, as convenient shorthand notation, we will also refer to the $a/f_a$ combination as $\theta$ so that 
 \be
  \theta = \frac{a_{\rm ph}}{f_a} +\theta_{\rm ind}.
  \ee

\subsection{Theta term: $\bar g_{aNN}(\theta)$}

Our strategy in computing the axion-nucleon coupling will be to start from $ {\cal L} = \theta \frac{\al_s}{8\pi} G \tilde{G}$ and evaluate the 
 order $\theta^2$ contribution to the nucleon mass. Then, expanding the corresponding terms to first order in the physical axion mass and $\theta_{\rm ind}$, one obtains the corresponding value of $\bar g_{aNN}$,
 \begin{eqnarray}
\left.\frac{1}{2}\frac{d^2m_N(\theta)}{d\theta^2}\right|_{\theta=0}\theta^2 \to  \left.
\frac{d^2m_N(\theta)}{d\theta^2}\right|_{\theta=0}\theta_{\rm ind} \times \frac{a_{\rm ph}}{f_a}.
 \end{eqnarray}
 Therefore, as is well known, the problem reduces to finding the quadratic 
 terms in the dependence of the nucleon masses on $\theta$ (see {\em e.g.} Ref. \cite{Lee:2020tmi}).
 We include the corresponding result here for completeness. 
 
Taking the initial axion Lagrangian (\ref{aGG}), we perform the chiral rotation 
\be
 q \rightarrow e^{i\frac{\theta_q}{2}\gamma_5} q, \qquad q=u,d,
\ee
with $\theta_q = \frac{m_*}{m_q}\theta$ (so that $\theta = \theta_u + \theta_d$). This removes the $\theta$ coupling to  the $G\tilde{G}$ term, while transforming the quark mass terms to  
\begin{align}
 {\cal L}_4 &= -m_u u\bar{u} - m_d d\bar{d} + \frac{1}{2} \theta^2 m_* \left( \frac{u\bar{u} + d\bar{d}}{2}\right) \nonumber\\
      & \quad  + \frac{1}{2} \theta^2 m_* \frac{m_d - m_u}{m_d+m_u}\left( \frac{u\bar{u} - d\bar{d}}{2}\right) \nonumber\\
  & \quad - m_*\theta (\bar{u} i\gamma_5 u + \bar{d}i\gamma_5 d) + \cdots \label{L4}
\end{align}
Substituting $\theta = \theta_{\rm ind} +a_{\rm ph}/f_a$, we note that the third term in the first line leads to the axion mass, on expanding $\theta$. If instead, we isolate the terms linear in $\theta_{\rm ind}$ and retain only the isoscalar part, we obtain
\be
{\cal L}_\theta =   \frac{1}{2}\theta_{\rm ind} m_* \frac{a_{\rm ph}}{f_a} (\bar{u} u + \bar{d}d) - m_*\theta(\bar{u} i\gamma_5 u + \bar{d}i\gamma_5 d), \label{Lm}
\ee
which, in addition to the pseudoscalar axion couplings in (\ref{La}), also includes an induced scalar coupling. 
It then follows that the isoscalar axion-nucleon coupling is determined by the nucleon sigma term \cite{MW},
\begin{align}
  \bar g_{aNN} \times f_a =   \frac{m_*\theta_{\rm ind}}{m_u+m_d} \si_{\pi N}
 \sim 1.5\times 10^{-9}\, {\rm MeV} \times\frac{\theta_{\rm ind}}{10^{-10}}, \label{aNN}
\end{align}
where $\sigma_{\pi N} \equiv (m_u+m_d)\left\langle N| \bar{u} u + \bar{d}d |N\right\rangle/2 \sim 40-50$~MeV, and for numerical estimates we have inserted the current bound on $\theta$ from measurements of hadronic EDMs \cite{review}. Note that the isovector coupling is also induced, but it is 
suppressed since $(m_u-m_d)\left\langle p| \bar{u} u - \bar{d}d |p\right\rangle/(2\sigma_{\pi N}) \sim O(10^{-2})$.

\subsection{Quark EDMs: $\bar g_{aNN}(d_q)$}

A variety of higher-dimensional sources of $CP$-violation can induce long-range axion forces \cite{P97}. As one example, we will consider the EDMs of quarks \cite{review} as part of ${\cal L}_{{\rm dim}\geq 5}$,
\be
 {\cal L}_{\rm EDM} = - \sum_q \frac{i}{2} {d}_q \bar{q} F_{\mu\nu}\sigma_{\mu\nu} \gamma_5 q. \label{edm}
\ee
The chiral rotation, $q \rightarrow e^{i\frac{\theta_q}{2}\gamma_5} q$, used to bring the axion coupling to the form (\ref{L4}), now induces the following terms
\begin{align}
 {\cal L}_{\rm EDM} &\rightarrow  \sum_{q=u,d,s} \frac{1}{2} m_* \theta F_{\mu\nu} \frac{d_q}{m_q} \bar{q} \si^{\mu\nu} q,
\end{align}
where we can simply identify $\theta = a_{\rm ph}/f_a$ in this case. 

Using the definition for the EM-polarizability of the QCD vacuum 
\begin{equation}
\langle 0| \bar q \sigma_{\mu\nu} q|0\rangle_F \equiv F_{\mu\nu} \times {e_q \chi} \langle \bar qq\rangle, 
\end{equation}
and assuming $SU(3)$ flavour invariance, we arrive at the following coupling of the axion field to the square of the 
EM field strength,
\begin{equation}
\frac{a}{2f_a} (eF_{\mu\nu})^2\times \chi \langle \bar qq\rangle \times \sum_{q=u,d,s} \frac{d_q Q_q m_*  }{m_q e},
\end{equation} 
where $Q_{u,d,s} = 2/3, -1/3, -1/3$. 

The nonperturbative succeptibility parameter $\chi$  \cite{Ioffe:1983ju,Vainshtein:2002nv,review}, is given by the following analytic expression derived by Vainshtein with the assumption of pion pole dominance, 
\begin{equation}
\chi = - \frac{3}{4\pi^2 F_\pi^2},
\end{equation}
which slightly exceeds earlier numerical estimates of $\chi \simeq 5$\,GeV$^{-2}$ \cite{Belyaev:1984ic}. 

In a relatively large nucleus, the operator $(eF_{\mu\nu})^2$ can be treated almost classically, a la Weisskopf, and approximated as 
\begin{equation}
 \int (eF_{\mu\nu})^2 d^3x \simeq -\frac{48\pi}{5}\frac{(Z\alpha)^2}{R_N},
\end{equation}
where a constant charge density sphere model 
is used for the nucleus, with radius $R_N\simeq 1.2\,{\rm fm} \times A^{1/3}$. Taking this estimate {\em per nucleon} we arrive at the effective $\bar g_{aNN}$ constant induced by the quark EDMs:
\begin{equation}
 \bar g_{aNN}^{\rm eff} \times f_a \simeq \frac{18(Z\alpha)^2 \langle \bar qq\rangle }{5\pi A F_\pi^2 R_N } \times 
\sum \frac{d_q Q_q m_*  }{m_q e}.
\end{equation}
Inserting the numerical factors, $\langle \bar qq\rangle  \sim (250\,{\rm MeV})^3$, $F_\pi \simeq 93$\,MeV, and taking $Z\simeq A/2$, we obtain the following numerical estimate for typical quark EDM values close to the current bounds: 
\begin{equation}
\label{EDMresult}
\bar g_{aNN}^{\rm eff} \times f_a \simeq 5\times 10^{-14}\, {\rm MeV} \times 
\frac{\sum \frac{d_q Q_q m_*  }{m_q e} }{10^{-26}\,\rm cm},
\end{equation}
with $Z=50$  used for the estimate.

\subsection{Quark Chromo-EDMs: $\bar g_{aNN}(\tilde d_q)$}

We can also consider the color EDMs of quarks \cite{review} as a source for the long-range interaction,
\be
 {\cal L}_{\rm CEDM} = - \sum_q \frac{i}{2} \widetilde{d}_q \bar{q} G \sigma \gamma_5 q. \label{cedm}
\ee
Moreover, given the poorly known scalar $s$-quark operator matrix elements in nucleon states, 
for this subsection, we concentrate on $u$ and $d$ color EDMs. 

Similarly to the EDM sources considered above, the chiral rotation, $q \rightarrow e^{i\frac{\theta_q}{2}\gamma_5} q$ used to bring the axion coupling to the form (\ref{L4}), also induces a shift in the CEDM sources,
\begin{align}
\label{cedm1}
 {\cal L}_{\rm CEDM} &\rightarrow - \sum_q \frac{i}{2} \widetilde{d}_q \bar{q} G\sigma \gamma_5 q \nonumber\\
                                   & \qquad + \frac{1}{2} m_* \theta\left(\frac{\widetilde{d}_u}{m_u} \bar{u} G\si u + \frac{\widetilde{d}_d}{m_d} \bar{d} G\si d \right),
\end{align}
where $\theta =  \theta_u + \theta_d$. 
The second line again leads to a direct contribution to the scalar $aNN$ coupling. However, the CEDM sources also induce a linear term in the axion potential so that, on relaxation to the minimum, a finite value of $\theta_{\rm ind}$ remains. This leads to a further contribution to the scalar axion coupling through (\ref{aNN}). 

To determine these contributions, we first note that Lagrangian (\ref{cedm1}) contains terms that can mediate transitions between the vacuum and pseudoscalar mesons that are light in the chiral limit: $\pi^0$ and octet $\eta$. One can account for these diagrams either explicitly  by combining the tadpole vacuum-to-pseudoscalar vertex with the $\pi\pi NN$ rescattering terms (see {\em e.g.} \cite{review}),
or equivalently, by chirally transforming (\ref{cedm1}) further, so that the tadpoles disappear. 
The parameters of such a rotation $\theta_q'$ are determined via
\be
 \left\langle 0 \left| - \sum_q \frac{i}{2} \widetilde{d}_q \bar{q} G\sigma \gamma_5 q - \sum_q \theta_q' m_q \bar{q}i\gamma_5 q \right| \pi^0\right\rangle = 0.
\ee
Requiring that $\theta_u' + \theta_d' = 0$ so that $\theta G\tilde G$ is not re-generated, one finds a correction $\theta_u' = \theta_d' \propto (\widetilde{d}_u - \widetilde{d_d})$ that multiplies only iso-triplet operators. This way we arrive at the following complete but somewhat lengthy expression that includes dimension 4 and 5 operators and where we retain only the terms that will contribute to $\bar g_{aNN}$:
\begin{eqnarray}
{\cal L}= \frac{1}{2} \theta^2 m_* \left( \frac{u\bar{u} + d\bar{d}}{2}\right)+\frac{1}{2}\theta^2m_*\frac{m_d-m_u}{m_d+m_u}\times 
\frac{\bar uu -\bar dd}{2}\nonumber\\
+\theta m_* m_0^2 \frac{\tilde d_u - \tilde d_d}{m_u+m_d}\times \frac{\bar uu -\bar dd}{2}~~~~~~~~~~~~~~~~~~\nonumber\\
+\frac12 \theta m_*\left(\frac{\tilde d_u}{m_u} 
\bar{u} G\si u +\frac{\tilde d_d}{m_d}\bar{d} G\si d \right),~~~~~~~~~~~~
\label{cedm2}
\end{eqnarray}
where $m_0^2 = \frac{\left< 0| \bar{q} G\sigma q|0\right>}{\left<0|\bar{q} q|0\right>} \sim 0.8$~GeV$^2$.

For most applications, the iso-singlet coupling is the most relevant. Thus, 
using 
\begin{align}
\frac{\tilde d_u}{m_u} \bar{u} G\si u +\frac{\tilde d_d}{m_d}\bar{d} G\si d
&= \left(\frac{\tilde d_u}{m_u} + \frac{\tilde d_d}{m_d} \right) \left(\frac{\bar{u} G\si u +\bar{d} G\si d}{2}\right)\nonumber\\
&\!\!\!\!\!\!\!\!\!\!\!\!\!\!\!\!\!\!\!\!\!\!\!\!\!\!\!+\left(\frac{\tilde d_u}{m_u} - \frac{\tilde d_d}{m_d} \right) 
\left(\frac{\bar{u} G\si u -\bar{d} G\si d}{2}\right),
\end{align}
we project Eq.(\ref{cedm2}) onto the iso-singlet part which then takes the form
\begin{align}
 \left.{\cal L}\right|_{\rm iso-singlet} &= \frac{1}{2} \theta^2 m_* \left( \frac{u\bar{u} + d\bar{d}}{2}\right) \nonumber\\
  &\!\!\!\!\!\!\!\!\!\!\!\!\!\!\!\!\!\!\!\!\!\!\!\!\!\!\!+ \frac{1}{2} m_* \theta \left( \frac{\widetilde{d}_u}{m_u} + \frac{\widetilde{d}_d}{m_d}\right) \left(\frac{\bar{u} G\si u + \bar{d} G\si d}{2} \right).
\end{align}
This Lagrangian accounts for both direct and induced contributions to the scalar axion-nucleon coupling. Extremizing in $a_{\rm ph}$ to find the minimum of the axion potential, we obtain 
\be
 \theta_{\rm ind} = - \frac{m_0^2}{2} \sum_q \frac{\widetilde{d}_q}{m_q} .
\ee
The resulting contribution to the scalar axion nucleon coupling is given by (\ref{aNN}).  Finally, on expanding $\theta^2$ terms and retaining the linear terms in $a_{\rm ph}$, we obtain the nucleon couplings as follows
\begin{align}
 \bar g_{aNN} &\times f_a = \frac{1}{2} m_* \left( \frac{\widetilde{d}_u}{m_u}+ \frac{\widetilde{d}_d}{m_d}\right) \nonumber\\
  &  \times \left\langle N \left| \frac{\bar{u} G\si u + \bar{d} G\si d}{2}  - m_0^2 \frac{ u\bar{u} + d\bar{d}}{2}  \right| N \right\rangle.
\end{align}
This expression clearly satisfies the tadpole cancelation requirement, namely that the result should vanish on replacing the nucleon state with the vacuum. In this case the direct and induced contributions precisely cancel. The remaining matrix element is the same as the one encountered in computing the $CP$-odd pion-nucleon couplings induced by CEDMs \cite{P01,review}. To appropriately translate the results of that work, we introduce singlet and triplet operators ${\cal H}$, 
\begin{equation}
\label{Hi}
{\cal H}^{(i)}= \frac{\bar{u} G\si u \pm \bar{d} G\si d}{2}  - m_0^2 \frac{ u\bar{u} \pm d\bar{d}}{2} ,
\end{equation}
where the choice of plus sign corresponds to isosinglet ${\cal H}^{(0)}$ and the minus sign to isovector ${\cal H}^{(1)}$. 
According to \cite{P01},
\begin{equation}
\label{Hhierarchy}
\left|\langle N | {\cal H}^{(0)}  | N \rangle \right|  \gg \left|\langle N | {\cal H}^{(1)}  | N \rangle\right|,
\end{equation}
where the singlet matrix element is estimated as,
\begin{equation}
\label{Hmatel}
\langle N | {\cal H}^{(0)}  | N \rangle  \simeq 0.6\,{\rm GeV}^2.
\end{equation}
With this result,
we obtain the iso-singlet coupling, 
\be
\label{CEDMresult}
\bar g_{aNN}(\widetilde{d_q}) \times f_a \simeq 1.5\times 10^{-10}\,{\rm MeV}
 \times  \left(\frac{\frac{m_*\widetilde{d}_u}{m_u} + \frac{m_*\widetilde{d}_d}{m_d}}{10^{-26} \,{\rm cm}}\right).
\ee
We have normalized the CEDM sources according to the generic limit from the neutron and $^{199}$Hg EDM constraints \cite{Hgnew,n}, in the absence of tuning. The ongoing lattice QCD effort to calculate matrix elements relevant for the $CP$-odd EDM-related observables may eventually improve on the estimate of Ref.~\cite{P01} regarding the size of the $\langle {\cal H} \rangle$ nucleon matrix elements, which will then also improve the accuracy of (\ref{CEDMresult}).

We note, in passing, that a much larger numerical coefficient for our CEDM treatment, Eq.~(\ref{CEDMresult}), compared to (\ref{EDMresult}), strongly suggests that a more important mechanism for inducing $\bar g_{aNN}$ from quark EDMs is through their radiative  mixing with CEDM, if all quantities in ${\cal L}_{dim\geq 5}$ are induced at the weak scale and/or above. The same could apply to the $CP$-odd Weinberg operator as well, $GG\tilde G$ that radiatively mixes with the flavour-singlet combination of quark color EDMs. Direct contribution of $GG\tilde G$ to $\bar g_{aNN}$ at low energy are likely to be suppressed.

\subsection{CKM phase: $\bar g_{aNN}(\delta)$}

The presence of $CP$-violation in the quark mixing matrix is a well-studied topic of the SM, both experimentally and theoretically. 
A natural question is how large the axion-mediated  long range force between nucleons would be, if it were sourced solely by the CKM phase. Assuming that it is flavor-diagonal, the result will be proportional to the reduced Jarlskog invariant $J\sim 10^{-5}$ \cite{jarlskog}. Since flavour diagonal $CP$-odd effects are necessarily of second order in the weak interactions, we anticipate that the coupling will be $\propto J G_F^2$. Arguments based on naive dimensional analysis (NDA) and chiral effective field theory have been used to argue that that scale of the coupling is \cite{GR},
\begin{align}
\label{CKMest}
 \bar g_{aNN}(\delta)|_{\rm NDA} \times f_a &\sim m_*JG_F^2 F_\pi^4  \sim 10^{-18}\,{\rm MeV},
\end{align}
which is highly suppressed. This suppression is fully expected due to the fourth power of the ratio of the QCD scale to the weak scale, and the smallness of the CKM angles. To a certain extent,  this estimate is very uncertain, due to the high power of hadronic scale in the numerator.  If we consider Eq.~(\ref{CKMest}) to be induced by loops, then  one may include numerically small factors, but then some of the hadronic scales may be traded for the charm quark mass, $F_\pi^2 \to m_c^2$. Thus, in reality $\bar g_{aNN}(\delta)$ may be significantly smaller or larger than the estimate (\ref{CKMest}) would suggest. Our goal in this section is to test the estimate above with explicit calculations.

We note that short distance radiative effects, relative to the QCD scale, do induce 
$d_d$ and $\tilde d_d$ operators at $G_F^2 m_c^2$ order, so that effectively two powers of $F_\pi$ in (\ref{CKMest}) are replaced with $m_c^2$. Unfortunately, such contributions come with the associated numerical suppression of three loops, and the corresponding Wilson coefficients are expected to be small: we estimate $\tilde d_d(\delta)$ as $g_s \tilde d_d \sim (g_s/(e/3)) d_d$, 
while $d_d$ was calculated in \cite{Czarnecki:1997bu}, and found to be 
$d_d \sim 0.7 \times 10^{-34} \,{\rm cm} \times (m_d/(10\,{\rm MeV}))$. Using this together with the CEDM result (\ref{CEDMresult}), we arrive at the following estimate of the short distance contributions to $\bar g_{aNN}$, 
\begin{equation}
\label{CKMestSD}
\left.  \bar g_{aNN}\right|_{\rm short\, dist.} \times f_a \sim 10^{-18}\,{\rm MeV},
\end{equation}
which interestingly is close to the NDA estimate (\ref{CKMest}).

Long-distance effects dominate $d_n(\delta)$ and contributions to $\bar g_{\pi NN}$ \cite{kz,Gavela,mu,dhm}, and since very similar physics is involved, such effects may well dominate $\bar g_{aNN}$. The goal for the rest of this section is to provide an estimate of the long distance contributions to $\bar g_{a NN}$  by describing it as a combination of two strangeness-changing transitions, $\Delta S = \pm 1$. As is well known, long-distance effects 
are crucially important for non-leptonic $|\Delta S| = 1$ effects in kaons and hyperons. 
In particular, studies of hyperon physics can be employed for a ``data-driven" estimate of $\bar g_{aNN}(\delta)$. Specifically, we will follow the approach where one of the weak $\Delta S = \pm 1$ vertices is identified with the 
nucleon-hyperon mixing term, and in that sense our calculation is closely related to approach of Ref.~\cite{Gavela}.

\begin{figure}[t!]
\centerline{\includegraphics[viewport=40 560 600 820, clip=true, scale=0.33]{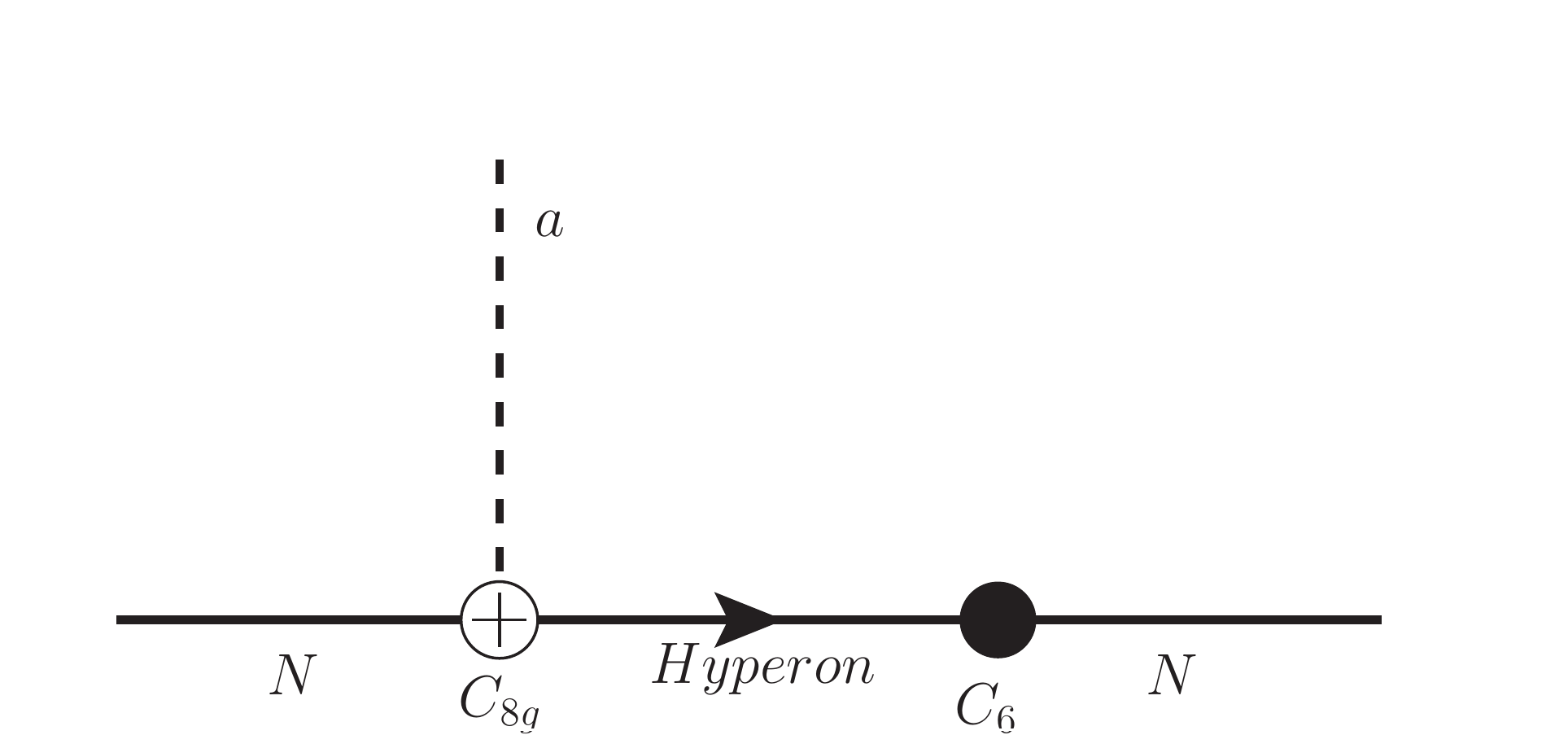}}
\caption{The pole diagrams contributing to the induced axion-nucleon coupling in the presence of the Standard Model CKM phase. See the text for further details of the vertices.} 
\label{fig:pole}
\end{figure}

To proceed, we start with the $\De S=1$ effective Lagrangian, given by
\be
 {\cal L}^{\De S=1}_{\rm eff} = \frac{G_F}{\sqrt{2}} V_{us}^* V_{ud} \sum_{i=3,\ldots,6,8g} C_i(\mu) {\cal O}_i + h.c.
\ee
where $V_{qq'}$ are CKM matrix elements. The Wilson coefficients $C_i(\mu)$ contain the dependence on the CKM phase.
A diagram of interest for $\bar g_{aNN}$ is pictured in Fig.~\ref{fig:pole}, 
where the black dot refers to the $CP$-even hyperon-nucleon vertex
dominated by the long-distance $\Delta S=1$ operators. There are strong indications that  this transition is dominated by the 
QCD-evolution-enhanced penguin operator ${\cal O}_6$ \cite{SVZ},
\begin{align}
{\cal O}_6  &= \bar{s}_\al \gamma_\mu (1- \gamma_5) d_\beta \sum_q \bar{q}_\beta \gamma^\mu (1+\gamma_5) q_\al, 
\end{align}
and its naive extrapolation to very low hadronic normalization scales may provide a satisfactory description of the famous 
empirical $\Delta I =1/2$ rule. Assuming the dominance of lowest order chiral perturbation theory leads to the following matrix elements for the $\Delta S=\pm 1$ transition amplitudes \cite{Bijnens:1985kj}: 
\begin{equation}
\label{Bchiral}
{\cal L}_{\Delta S = \pm1} = -a_W {\rm Tr}(\bar B\{ h,B\})- b_W {\rm Tr}(\bar B[ h,B]) +h.c.,
\end{equation}
where $B$ is the standard baryon matrix in flavour space, $h$ is a $3\times3$ matrix that has a single non-vanishing matrix element $h_{23} = 1$, and the weak amplitudes $a_W,\,b_W$ are determined from the fit to the $s$- and $p$-wave amplitudes of the hyperon decays. A tree-level fit to data gives \cite{Bijnens:1985kj}
\begin{eqnarray}
a_W = \tilde a_W \times \sqrt{2}G_FF_\pi m_{\pi^+}^2;~ b_W= \tilde b_W \times\sqrt{2}G_FF_\pi m_{\pi^+}^2, \nonumber\\
\tilde a_W = 0.56; ~ \tilde b_W = -1.42.~~~~~~~~~~~
\label{Wampl}
\end{eqnarray}
In these expressions, $m_{\pi^+}$ should be understood as simply the numerical scale of $139.6$ MeV, which is independent of the values of light quark masses rather than a theoretical quantity that tends to zero in the chiral limit. Given the fact that the external particles are either $n$ or $p$, one can truncate Eq.~(\ref{Bchiral}) to 
\begin{eqnarray}
{\cal L}_{\Delta S = \pm1} = -a_W \left( -\frac16\bar n (\sqrt{6}\Lambda +3\sqrt{2}\Sigma^0)+\bar p \Sigma^+\right)
\nonumber\\
-b_W\left( -\frac12\bar n (\sqrt{6}\Lambda -\sqrt{2}\Sigma^0)-\bar p \Sigma^+\right)+h.c.~~~
\end{eqnarray}

In Fig.~\ref{fig:pole}, the cross corresponds to a $CP$-violating vertex with a non-derivative coupling to the axion, and therefore it has to be chirality-flipping at the quark level. 
Among all possible $\Delta S=1$  operators of lowest dimension, only one such operator violates chirality, and it is the chromo-magnetic $s-d$ dipole operator ${\cal O}_{8g}$,
\begin{align}
\label{8g}
{\cal O}_{8g} &= \frac{1}{8\pi^2} m_s \bar{s} (G  \si) (1-\gamma_5)d.
\end{align}

Examining the Wilson coefficients, one finds at leading order an enhanced contribution of the top quark to 
$C_{8g}$, and as a consequence, a much larger imaginary part for the corresponding coefficient: 
\be
\left| \frac{{\rm Im}(C_{6})}{{\rm Re}(C_6)} \right| \simeq A^2\lambda^4 \et\quad  \ll \quad \left| \frac{{\rm Im}(C_{8g})}{{\rm Re}(C_{8g})} \right| \simeq  \et,
\ee
where $\lambda \sim 0.23$, $A\sim 0.79$ and $\eta\sim0.36$ are Wolfenstein parameters in the CKM matrix, and the leading $CP$-violating contributions arise via top loops. This justifies taking the $a_W$ and $b_W$ coefficients to be real. 
Next-to-leading order corrections are not negligible \cite{Bijnens:1985kj}, but maintain this hierarchy, which  allows us to focus on the chirality violating dipole operator ${\cal O}_{8g}$ as the primary source of $CP$-violation.

At the next step, we can include axion dependence via the $m_{d,s} \to m_{d,s} + im_*\theta \gamma_5$ substitution. 
It turns out that the axion-dependence of  ${\cal O}_{8g}$ in the limit of exact $SU(3)$ flavour invariance takes the following form:
\begin{equation}
{\cal O}_{8g} \to {\cal O}_{8g} + \frac{1}{4\pi^2} m_*\theta \bar{s} (G  \si) i\gamma_5d.
\end{equation}
One should note that the operator $\sim \theta \bar s (G\si) d$ cancels in this limit exactly, which also means that the 
$a_{\rm ph} \bar BB$ vertex is not generated. We note, however, that there is no profound reason why this 
operator should be absent, and speculate that after accounting for unspecified $SU(3)$ breaking effects one would 
expect non-vanishing $\theta \bar s (G\si) d$ terms. To this end, we introduce a phenomenological parameter $\kappa$ 
that accounts for $SU(3)$ violation (probably at the level of $|\kappa|\propto 0.2$), and write the axion-quark dipole operator in the following form: 
\begin{align}
\label{Wdipole}
&{\cal L}_{\Delta S =\pm 1, a} = \kappa \theta m_*  \left( \bar s(G\si)d - m_0^2 \bar sd  +h.c.\right) ~~~~
\\\nonumber & \qquad\times
 \frac{G_F}{\sqrt{2}} \frac{{\rm Im}[V_{td}V_{ts}^*]}{4\pi^2}\left(\frac{x_t^3-5x_t^2-2x_t}{8(x_t-1)^3}+
 \frac{3x_t^2\log x_t}{4(x_t-1)^4}
 \right).
\end{align}
where $x_t=m_t^2/m_W^2$.
The numerical value of the second line in this expression, which we will call $G_{\rm loop}$  is $ \frac{G_F}{\sqrt{2}}\times 3.3\times 10^{-7}$. (Part of this suppression comes from the expression in parentheses, that turns out to be rather small, $\sim 0.1$.)

Notice that there is an additional subtraction of $m_0^2 \bar sd $ present in (\ref{Wdipole}). This subtraction can be justified more generally as follows. Notice that $\theta$ in (\ref{Wdipole}) is constant, and therefore is momentum-independent. Thus, any quark bilinear operator $\bar s O d$ where $O$ is some combination of gamma matrices and gluon fields should be modified due to the following argument: if $O$ is replaced with the unit operator, its physical effect must disappear as it would then represent a slight correction to the quark mass basis that does not have any real physical effect. Therefore, the expected modification to any such operator in the $SU(3)$ limit should be 
\begin{equation}
\bar s O d \to \bar s O d- \bar s d\times \frac{\langle 0| \bar q Oq |0\rangle }{\langle 0| \bar q q |0\rangle }.
\end{equation}
In chiral perturbation theory such subtractions typically arise after the inclusion of tadpole and mass insertions into the pseudoscalar meson lines in diagrams.

The operator in (\ref{Wdipole}) can be translated to the $a\bar BB$ vertex using the same matrix elements we encountered in the evaluation of $\bar g_{aNN}(\tilde d_q)$. Notice that the operator in parentheses from the first line of 
Eq.~(\ref{Wdipole}) can be viewed as an $SU(3)$ generalization of (\ref{Hi}), ${\cal H}_{sd} = \frac{1}{2} ( {\cal H}^6 + i {\cal H}^7)$, where 
\be
 {\cal H}^a = \bar{q} G \si \lambda^a q - m_0^2 \bar{q}\lambda^a q.
\ee
The required vertices then follow from $SU(3)$ flavour symmetry, with two unknown parameters
\be
 \langle \bar{B} | {\cal H}^a | B \rangle = -d_1 {\rm Tr} (\bar{B} \lambda^a B) - d_2 {\rm Tr}(\bar{B} B \lambda^a). \label{su3}
\ee
It follows that 
\begin{align}
 \langle B | {\cal H}_{sd} | B \rangle &= d_1 \left(  \sqrt{\frac{2}{3}} \bar{n} \La\right) \nonumber \\
     & - d_2 \left ( \bar{n} \left( \frac{1}{\sqrt{6}} \La - \frac{1}{\sqrt{2}} \Si^0\right) + \bar{p} \Si^+\right) +\cdots.
\end{align}
We can determine $d_1$ and $d_2$ via considering matrix elements over the proton. From (\ref{su3}) we have $\langle p| {\cal H}^3 |p\rangle = d_1 \bar{p} p$, and $\langle p| {\cal H}^8 |p\rangle = \frac{1}{\sqrt{3}} (d_1-2d_2) \bar{p} p$. If we define the diagonal flavour operators as ${\cal H}_i = \bar{q}_i G \si  q_i - m_0^2 \bar{q}_i q_i$, and take guidance from lattice and QCD sum rules estimates of similar operators with the same chiral structure, implying $\langle p | {\cal H}_s | p\rangle , \langle p | {\cal H}_u - {\cal H}_d | p\rangle \ll \langle p | {\cal H}_u + {\cal H}_d | p\rangle$, then we find that $d_1$ can be neglected while $d_2$ is given by (\ref{Hmatel}),
\begin{align}
 d_ 2  \sim\frac{1}{2} \langle p | {\cal H}_u + {\cal H}_d | p \rangle
       \sim 0.6\, {\rm GeV}^2.
\end{align}

\begin{figure}[t!]
\centerline{\includegraphics[viewport=40 400 600 780, clip=true, scale=0.33]{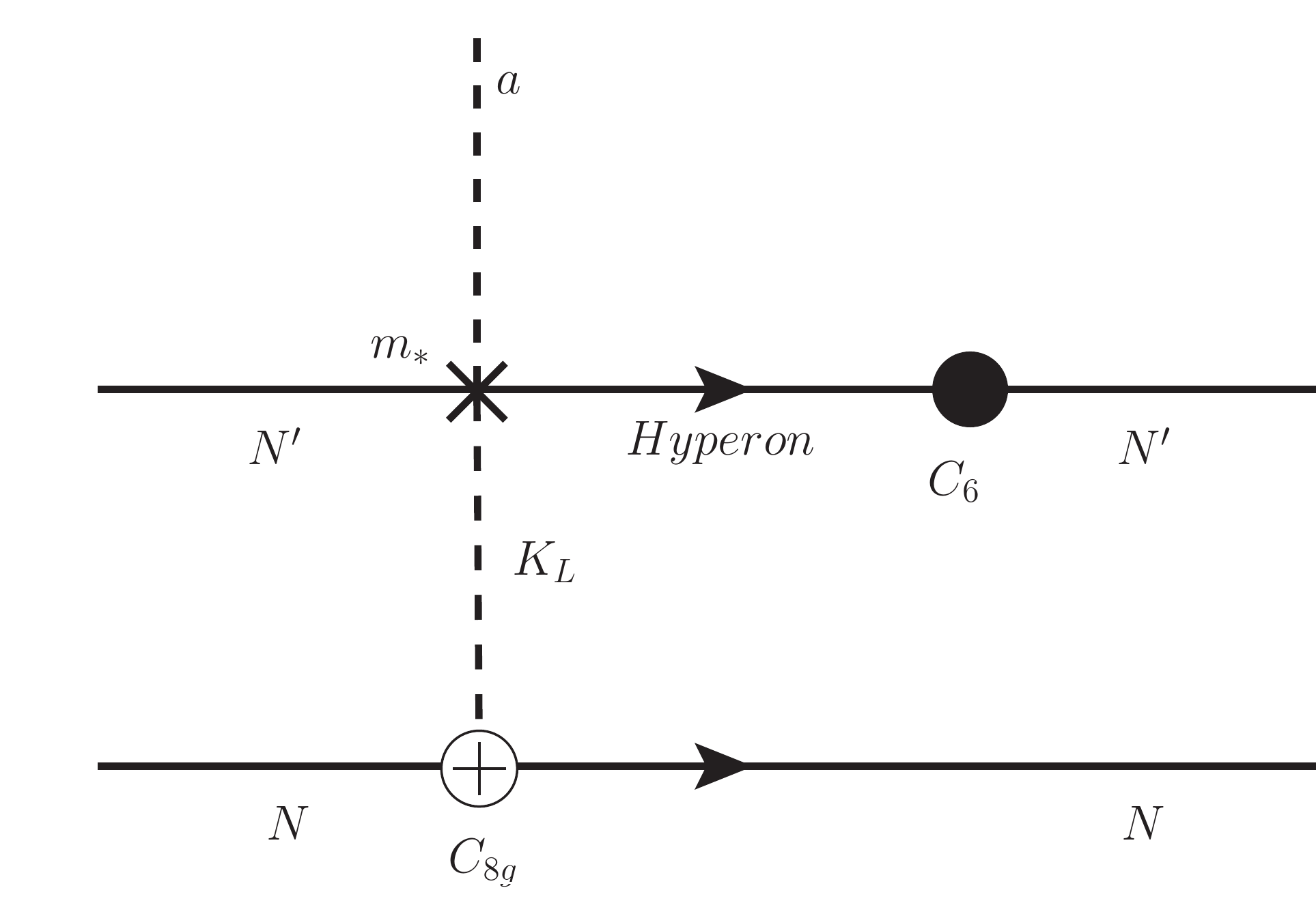}}
\caption{The diagrams contributing to the mean-field {\it equivalent} $aNN$ coupling in the presence of the Standard Model CKM phase. See the text for further details of the vertices.} 
\label{fig:aNN}
\end{figure}

Combining all the pieces together in the diagrams of Fig.~\ref{fig:pole}, we take into account $\Lambda$, $\Sigma^0$ and $\Sigma^+$ pole contributions.  We finally obtain the following estimate,
\begin{align}
&{\cal L}_{aNN} \sim \frac{a}{f_a}\times \kappa m_* d_2  G_F^2 F_\pi m_{\pi^+}^2\times 3.3\times 10^{-7} \times \\
&\left( \frac{\bar nn({-}\frac{\tilde b_W}{2}{-}\frac{\tilde a_W}{6}) }{m_n-m_\Lambda}  +
\frac{\bar nn({-}\frac{\tilde b_W}{2}{+}\frac{\tilde a_W}{2}) }{m_n-m_{\Sigma^0}} +
\frac{\bar pp({-}\tilde b_W{+}\tilde a_W) }{m_p-m_{\Sigma^+}}  \right). \nonumber
\end{align}
Numerically evaluating this expression and extracting the dominant isoscalar component, we obtain
\begin{equation}
\label{fakappa}
\bar g_{aNN}(\delta) \times f_a \sim 1 \times 10^{-18}\,{\rm MeV} \times \kappa,
\end{equation}
which is close to but somewhat subdominant, on account of $\kappa$ being small, 
to the naive estimate (\ref{CKMest}) and to the short-distance contribution (\ref{CKMestSD}). 

Finally, we would like to discuss the possibility of generating an {\em equivalent} $\bar g_{aNN}$ coupling via meson exchange inside a large nucleus, with $CP$ violation sourced again by $\delta_{CKM}$. We will calculate contributions to operators of the form $ a(\bar NN)(\bar NN)$, and evaluate the effective $aNN$ coupling using the mean free field approximation. 
A representative diagram is shown in Fig.~\ref{fig:aNN}. The kaon exchange that mediates $\Delta S =\pm 1$ transitions can be thought of as a contact interaction, as $m_K$ is larger than the typical nucleon momenta. Therefore, the result of kaon exchange can be expressed via a single coupling,
\begin{equation}
\label{4N}
{\cal L}_{4N} = - g_{4N} \times a_{\rm ph} (\bar NN)(\bar NN).
\end{equation}
Inside a large nucleus, there is an approximately constant number 
density 
\begin{equation}
n_N \simeq A/V_N \simeq 0.15\, ({\rm fm})^3 \simeq (106\,{\rm MeV})^3.
\end{equation}
The energy scale associated with $n_N$  is not small compared to other 
low-energy QCD parameters such as $F_\pi$ and $m_\pi$. 
The interaction (\ref{4N}) gives the axion coupling to the whole nucleus as $g_{4N} n_N^2 V_N = g_{4N} n_N A$, and therefore the equivalent axion-nucleon coupling is given by
\begin{equation}
g_{aNN}^{\rm equiv} \simeq g_{4N} \times n_N.
\end{equation}

Next we evaluate the CKM phase contribution to $g_{4N}$. First, we evaluate the kaon coupling to the nucleon. 
The $CP$-odd  gluonic dipole operator, $O_{8g}$ (\ref{8g}) generates non-derivative $K_L \bar NN $ coupling, and the result of an explicit calculation leads to 
\begin{equation}
{\cal L}^{CP-{\rm odd}}_{KNN}= K_L(\bar pp + \bar nn) \times G_{\rm loop} \times\frac{m_s }{2 F_\pi} \langle N| {\cal H}^{(0)}|N\rangle ,
\end{equation}
where the matrix element is exactly the one encountered before in Eq. (\ref{Hmatel}).

The $CP$-even block of the diagram Fig.~\ref{fig:aNN} is given by the already familiar hyperon pole contributions. 
The new element in the calculation encountered here is the non-derivative $a_{\rm ph}K_L \bar NN$ vertex that 
is the direct analogue of the nucleon $\sigma$-term in the $\Delta S =\pm 1$ channel, and its size is dictated by the matrix element of the $(F_\pi f_a)^{-1}m_*(\bar ds +\bar sd)$ operator. From the point of view of $SU(3)$ flavour symmetry, this operator transforms as an octet, and therefore its matrix elements are given by exactly the same $SU(3)$ structures as those proportional to $m_s\lambda_8$ that lead to the $s$-quark mass-induced baryon mass splitting, Eq.~(\ref{su3}). Combining this vertex, the baryon poles, and the weak interaction induced transitions proportional to $a_W$ and $b_W$, Eq. \ref{Bchiral}, we end up with the following $CP$-conserving interaction of protons and neutrons with $a K_L$, 
\begin{equation}
{\cal L}^{CP-{\rm even}}_{KNN}= (2.0 \bar pp + 2.8 \bar nn) \frac{\theta K_L}{F_\pi}
\times \frac{2m_*}{m_s} \sqrt{2}G_FF_\pi m_{\pi^+}^2.
\end{equation}
Notice that $m_s^{-1}$ in this expression is a remnant of the baryon pole, while $m_{\pi^+}$ is again simply the numerical value of the pion mass that remains fixed in the chiral limit. For a large nucleus, $2.0 \bar pp + 2.8 \bar nn$ can be approximated as $\simeq 2.5 \bar NN$. Finally, putting together  ${\cal L}^{CP-{\rm even}}_{KNN}$ and ${\cal L}^{CP-{\rm odd}}_{KNN}$ and integrating out the $K_L$ field, we obtain $g_{4N}$, and the equivalent $\bar g_{aNN}$ coupling
in the following approximate form,
\begin{equation}
 g_{aNN}^{\rm equiv} \times f_a \simeq 5\times 10^{-7}\times {\rm GeV}^2 \times  \frac{G_F^2m_{\pi^+}^2n_Nm_*}{m_K^2F_\pi }.
\end{equation}
Numerically, this corresponds to 
\begin{equation}
\label{gaCKMeq}
 g_{aNN}^{\rm equiv}(\delta) \times f_a \simeq  2\times 10^{-19}\,{\rm MeV} ,
\end{equation}
which is again remarkably close to the earlier estimates. The advantage of (\ref{gaCKMeq}) is that it does not have an indeterminate parameter $\kappa$ as in Eq.~(\ref{fakappa}), and it dominates in the chiral limit, as then $\kappa \to 0$. We also note that this contribution does not deviate too far from the crude NDA estimate (\ref{CKMest}), and is a factor of a few smaller than the estimate of the short-distance contribution (\ref{CKMestSD}).

\section{Discussion}

Ongoing experimental efforts to search for spin-mass and mass-mass couplings mediated by axion forces \cite{Raffelt:2012sp} 
motivate reconsideration of the $CP$-odd coherent $\bar g_{aNN}$ coupling. In this paper we have revisited the calculation of this coupling, concentrating on its generation by beyond the SM contributions coming from dimension $\leq 5$ operators, and on its baseline CKM phase-induced contribution. 

Our results show that the induced theta term contributing to $\bar g_{aNN}$ may only dominate in models where $CP$ violation arises at very low energy, through the generation of `tadpole' contributions linear in the axion field. In all scenarios with $CP$-violation originating at or above the electroweak scale, and parametrized by effective higher-dimensional operators, the induced $\theta$ term is only one contribution, inseparable from `direct' contributions. This is the case for CEDM sources, for example. The results for the CEDM-driven value for $\bar g_{aNN}$ depends on the poorly known matrix element of the ${\cal H}$ operators, estimated in \cite{P97}. These are the same matrix elements that determine the $CP$-odd  $\bar g_{\pi NN}$ effective couplings, and future progress in calculating these matrix elements on the lattice will also improve the quality of these estimates. In addition, we have calculated the direct effect of the quark EDMs on $\bar g_{aNN}$ but the result appears to be rather small, so that RG mixing of EDM$\to$CEDM is likely the dominant source. 

We have carefully considered the CKM-induced value of $\bar g_{aNN}$, finding that long-distance contributions (from separate $\Delta S =\pm 1$ transitions) contribute to $\bar g_{aNN} \times f_a$ at the level of $\sim {\rm few} \times 10^{-19}$\,MeV, which is close to both the short-distance contribution from CKM-induced $\tilde d_d$, 
and to the naive dimensional analysis estimate $\sim 10^{-18}$\,MeV. We also find that the composite nature of the nucleus provides an additional calculable source of an `equivalent' $\bar g_{aNN}$ coupling, mediated by $a(\bar NN)^2$ 
interactions in the mean field approximation. It is easy to see, however, that all these CKM estimates of $\bar g_{aNN}$ are of academic interest only given current experimental sensitivity. They are below the limits on new physics that saturates current EDM bounds by about eight-to-nine orders of magnitude.
One should also bear in mind the approximate nature of these long-distance estimates, and with so many different terms and interactions at play, further long-distance contributions may be found.

\begin{acknowledgments}
M.P. would like to thank J. Redondo for encouragement to revisit the CKM contribution to $\bar g_{aNN}$. 
The work of  S.O. and A.R. is supported 
in part by NSERC, Canada, and MP is supported in part by U.S. Department of Energy (Grant No. desc0011842).
S.O. also acknowledges financial support from the State Agency for Research of the Spanish Ministry of Science and Innovation through the ``Unit of Excellence Mar\'ia de Maeztu 2020-2023'' award to the Institute of Cosmos Sciences (CEX2019-000918-M) and from PID2019-105614GB-C21 and 2017-SGR-929 grants.
\end{acknowledgments}

\end{document}